\documentclass{stanfest}
\usepackage[utf8]{inputenc}
\usepackage{amsfonts}
\usepackage{amsmath}
\usepackage{amssymb}
\usepackage{mathrsfs}%cursive font
\usepackage{graphicx}
\usepackage{color}
\usepackage{float}
\usepackage{hyperref}
\newcommand{\rhod}{\langle \rho_\mathrm{d} \rangle}
\newcommand{\rhoc}{\langle \rho_\mathrm{c} \rangle}
\newcommand{\be}{\begin{equation}}
\newcommand{\ee}{\end{equation}}
\author{Ilya Mandel}[Monash,OzGrav]
\author{Ryosuke Hirai}[RIKEN,Monash,OzGrav]
\author{Lewis Picker}[Monash]
\affil[Monash]{School of Physics and Astronomy, Monash University, Clayton, VIC 3800, Australia}
\affil[OzGrav]{The ARC Centre of Excellence for Gravitational Wave Discovery -- OzGrav, Australia}
\affil[RIKEN]{Astrophysical Big Bang Laboratory (ABBL), Cluster for Pioneering Research, RIKEN, Wako, Saitama 351-0198, Japan}
\title{Common envelopes at StanFest}

\begin{document}

\maketitle

\begin{abstract}
We describe some of our group's recent work on common envelopes.  Our goal is to understand the onset and outcomes of dynamically unstable mass transfer, including the properties of the binaries left behind and the outflows during the common envelope stage.  We have also started thinking about light curves of common envelope events.  During a talk at StanFest, a meeting in honour of Stan Owocki's retirement held in Leuven in July, 2024, the first author reported on some of the results we recently obtained and briefly outlined future prospects.
\end{abstract}

\section{Common Envelope Timescales and Outcomes}
\label{sec:intro}

In homage to Stan Owocki's ability to quickly illustrate the key physics of a problem with simple order-of-magnitude calculations, we will begin with a back-of-the-envelope  introduction to the timescales and outcomes of common envelope events.  For our purposes, a common envelope refers to a companion spiralling in through the envelope of the donor, which is no longer in co-rotation, and losing orbital energy due to drag forces \citep[see the review][for a more sophisticated discussion]{Ivanova:2013}. 

Consider a donor of mass $M$ and radius $R$ and an inspiralling companion of mass $m$ and radius $r$.  We will assume $m \ll M$ for simplicity, though this is not strictly necessary and will only lead to order unity errors when $m \sim M$, and of course $r \ll R$.  The orbital energy is $E \sim G M m / R$.  The donor's average density (we will ignore factors of order unity throughout) is $\rhod \equiv M/R^3$; the companion's average density is $\rhoc \equiv m/r^3$; and $\rho$ is the envelope density at the current location of the companion.

The companion's Keplerian orbital velocity is $v = \sqrt{GM/R}$; we will assume that the donor is not rotating.  The dynamical timescale (orbital period) is $\tau_\mathrm{dyn} = R/v$.  The Bondi-Hoyle radius of the companion is $r_B \sim G m / v^2 = R (m/M)$ (we are assuming super-sonic motion here, $v > c_s$, where $c_s$ is the speed of sound).  If $r_B>r$, i.e., $R > (M/m) r$ (this is the case, e.g., for compact-object companions), Bondi-Hoyle drag dominates, and the effective cross-section is $\sigma \sim r_B^2 \sim R^2 (m/M)^2$.  On the other hand, if $r>r_B$, ram-pressure drag dominates, and the effective cross-section is $\sigma \sim r^2$.

The mass-accretion rate is $\dot{m} =C_A \rho v \sigma$, where $C_A$ is the dimensionless accretion coefficient.  The drag force is $F = C_D \rho v^2 \sigma$, where $C_D$ is the dimensionless drag coefficient.  The energy dissipation rate is then $\dot{E} = - C_D \rho v^3 \sigma$.   Numerical experiments show that, unlike $C_A$, the drag coefficient $C_D$ is almost always of order unity \citep[e.g.,][]{MacLeodRamirezRuiz:2015,De:2019}, so we will generally ignore it below.

The inspiral timescale is $\tau_{insp} \equiv E/|\dot{E}|$.  Thus,
\be
\frac{\tau_\mathrm{insp}}{\tau_\mathrm{dyn}} \sim \frac{G M m} {R^2 \rho v^2 \sigma} = \frac{m}{R \rho \sigma}.
\ee

In the Bondi-Hoyle regime, $\sigma \sim R^2 (m/M)^2$, so
\be
\frac{\tau_\mathrm{insp}}{\tau_\mathrm{dyn}} \sim \frac{M^2}{R^3 m \rho} = \frac{M}{m}\frac{\rhod}{\rho}.
\ee
As an example, consider the inspiral of a neutron star into the convective envelope of a red supergiant en route to forming a double neutron star system.  In this case, the envelope's density is not too far off from uniform, so $\rhod/\rho$ is perhaps only a few except in the outermost envelope layers, and $M/m$ is only a few as well -- so the inspiral will proceed over a few orbits, which matches the findings of numerical simulations \citep[e.g.,][]{Lau:2021}.

In the ram-pressure regime, $\sigma \sim r^2$, so
\be
\frac{\tau_\mathrm{insp}}{\tau_\mathrm{dyn}} \sim \frac{m}{R r^2 \rho} = \frac{r}{R}\frac{\rhoc}{\rho}.
\ee
As an example, consider the engulfment of a planet by a star with a radiative envelope.  In the outermost layers of the star, the density is very low, $\rho \ll \rhoc$, so the inspiral timescale is very long -- the orbit is almost circular.  However, the density rapidly increases inward.  Since $r \ll R$, long before the tidal disruption of the planet at $\rhoc \sim \rho$, the inspiral timescale drops below the dynamical timescale once $\rho \sim \rhoc (r/R)$.  At this point, the orbital motion stalls, and the planet transitions to a largely radial infall with a terminal velocity $v_\mathrm{term}$ determined by equating the gravitational acceleration to drag, $G M m / R^2 \sim r^2 v_\mathrm{term}^2 \rho$.  Of course, the planet may ablate even sooner: if we imagine that the momentum imparted by ram pressure drives off planetary material with a velocity equal to the escape velocity at the surface of the planet, $v_\mathrm{esc} \sim \sqrt{Gm/r}$, then the ablation mass loss rate is $\dot{m} \sim \rho v^2 r^2 / v_\mathrm{esc}$, so the mass ablation timescale is only
\be
\frac{\tau_\mathrm{ablat}}{\tau_\mathrm{dyn}} \sim \frac{m}{\dot{m} \tau_\mathrm{dyn}} \sim \sqrt{\frac{R}{r}\frac{m}{M}}\
\end{equation}
once the planet is moving inward at the terminal velocity.  The term under the square root is typically much less than unity, so we may expect the planet to be ablated well before it is tidally disrupted.
%the typical energy release by the time the planet moves to radius $x \ll R$ is $\Delta E \sim G M m / x$, and if this energy is primarily deposited into the planet rather than the star, it can ablate once the deposited energy exceeds the planet's binding energy, $\Delta E > G m^2 /r$; 
See \citet{Lau:2022engulf} for numerical simulations and further discussion of this regime.

Note that with the drag and accretion coefficients included, 
\be
\frac{dE}{E} = - \frac{C_D}{C_A} \frac{dm}{m},
\ee
so $E_0/E = (m/m_0)^{C_D/C_A}$.  E.g., for $C_D=C_A=1$, the energy decay timescale is the same as the mass growth timescale.  If we believe that neutron stars can spiral in by several orders of magnitude within a common envelope while barely changing their mass, it must be the case that $C_A \ll C_D$.  Wind tunnel simulations suggest that this is precisely the case when the density and pressure scale heights of the donor are small relative to the Bondi radius, with the excess momentum in the stream closer to the donor's core suppressing accretion \citep{MacLeodRamirezRuiz:2015}.  However, resolution in such simulations is an issue, with many orders of magnitude separating the Bondi radius and the accretion radius \citep{De:2019}, which means that we cannot self-consistently model whether there is a centrifugally supported barrier close to the accretor \citep{Murguia:2020} nor what kind of feedback \citep[perhaps via jets? see, e.g.,][]{Soker:2004} this accretion produces.

It's worth briefly commenting on our assumption of supersonic motion.  The sound speed is approximately $c_s^2 \sim P/\rho$.  Because the donor is in hydrostatic equilibrium, $dP/dr = G M \rho /R^2$.  Thus, on average, equating $dP/dr \sim P/R$, we would conclude that $c_s^2 \sim P/\rho \sim GM/R \sim v^2$ -- i.e., the sound speed is of the order of the Keplerian orbital velocity and our calculation is roughly accurate.  More precisely, the validity of this assumption depends on the details of the envelope structure.

What about the possibility of the binary surviving the common-envelope phase by ejecting the envelope?  Drag heats the envelope; the envelope expands in response to heating; as it does so, it may cool sufficiently for ionized helium and hydrogen to recombine, leading to an additional injection of recombining energy, perhaps causing the entire envelope to unbind \citep{Ivanova:2018,Lau:2022}.  Is there enough energy available to unbind the envelope?  The closest that the accretor can get to the donor without the binary merging is presumably a few times the core radius of the donor; the maximum change in orbital energy is then $\Delta E_\mathrm{orb}^\mathrm{max} \sim (G M_c m / R_c - G M m / R)$, where $M_c$ and $R_c$ are the donor's core mass (perhaps a few tens of percent of its total mass) and core radius.  For $R_c \ll R$, this is $\Delta E_\mathrm{orb}^\mathrm{max} \sim G M_c m / R_c$.  Meanwhile, the binding energy of the envelope is a quantity of order $G M^2 / R$, often parametrized as $E_\mathrm{bind} = G M (M-M_c) / \lambda / R$ with a parameter $\lambda$ that describes the envelope's structure \citep{deKool:1990}.  Thus, as long as $\Delta E_\mathrm{orb}^\mathrm{max} \gtrsim  E_\mathrm{bind}$, i.e., $R / R_c \gtrsim M/m/\lambda$, there should be enough energy available to unbind the envelope.  Since $R/R_c$ could be in the hundreds or even thousands for evolved stars, this seems fairly straightforward.  Alternatively, assuming that a fixed fraction of order unity of the orbital energy goes into unbinding the envelope, the binding energy and the donor core and companion masses can be converted into a final separation.  This orbital energy fraction is parametrized as $\alpha$ \citep{Webbink:1984} and often treated as a universal value in models of binary evolution.  However, a key assumption we will return to in a moment is whether the energy is deposited quickly enough that we can assume that it all goes into expansion, rather than being lost, e.g., to radiative cooling.  

In this model, the envelope is ejected on a dynamical timescale and with a dynamical velocity, i.e., a velocity of order the escape velocity from the donor $\sim \sqrt{GM/R}$; this seems to be born out by at least some 3-dimensional hydrodynamical simulations \citep{Lau:2021}.

\section{Common envelopes are not spherical cows in vacuum}

Of course, the real story is rather more complicated than the simple order-of-magnitude exposition above suggests (perhaps a good thing for those of us who want to remain employed in astrophysics).    

First, there is the question of when common envelopes occur, i.e., when mass transfer becomes dynamically unstable.  Two of the key ingredients here are how conservative mass transfer is and how much (specific) angular momentum is carried away by non-conservative mass transfer.  These two determine how the size of the donor's Roche lobe changes in response to mass loss.  Presumably, if the donor overflows the Roche lobe by ever larger amounts as mass transfer continues, the runaway process leads to a common-envelope phase \citep{Ivanova:2013}.

Why does mass transfer become non-conservative?  One can imagine at least three possibilities: accretor expansion, accretor spin-up, and Eddington-limited accretion onto a compact object.  

As mass is deposited on the accretor, mass settling leads to a release of gravitational energy.  If the accretor is unable to radiate this energy away, it expands.  If it expands enough to fill its own Roche lobe, mass transfer probably ceases to be conservative.  \citet{Lau:2024} recently explored the response of accretors to rapid mass gain. They confirmed \citep[e.g.,][]{Hurley:2002} that mass transfer does indeed become non-conservative when the rate of energy release through gravitational settling, $\sim G m \dot{M} / R_\mathrm{eff}$, significantly exceeds accretor luminosity $L$ (here, $R_\mathrm{eff}$ is the effective radius to which mass settles; a fit is provided by \citealt{Lau:2024}). However, they also discovered that the onset of rapid accretor expansion is a strong function of the accretor mass.  In particular, lower-mass accretors can avoid significant expansion even when accreting mass at a rate that is hundreds of times greater than their thermal timescale, allowing them to become ``hamstars" -- stars that behave like hamsters, saving material they can't promptly thermalise behind their puffy cheeks.  This is illustrated in Figure 1, adopted from \citep{Lau:2024}.

\begin{figure}[!t]
\begin{center}
\includegraphics[width=0.8\textwidth]{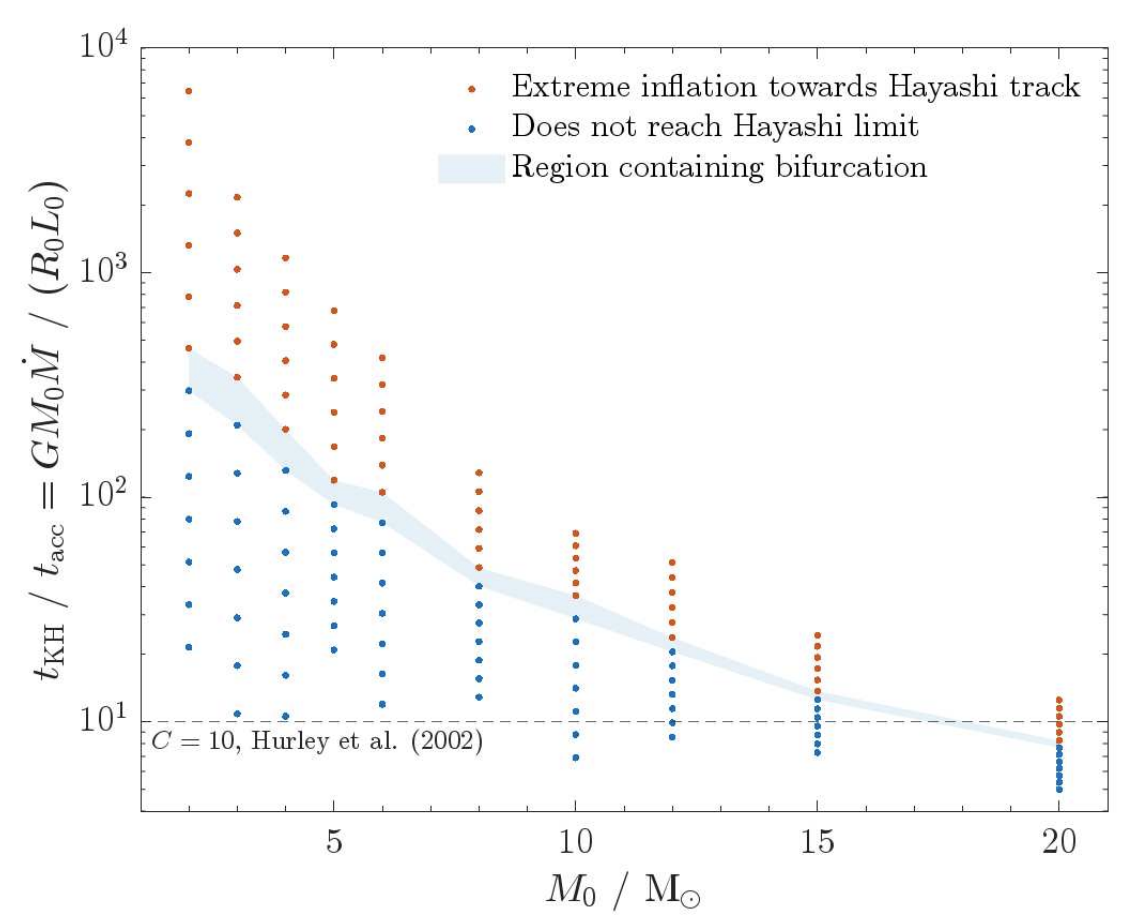}
\end{center}
\caption{Accretors that experience rapid inflation toward the Hayashi line (red) vs.~accretors that do not (blue).  The threshold between the two is expressed as a multiplier to the ratio of the accretor's thermal timescale to the accretion timescale (ordinate), but this threshold is a function of the accretor's initial mass (abscissa). Figure 1 from \citet{Lau:2024}.}
\end{figure}

Only a few percent of mass gain should be enough to bring the accretor to critical rotation.  This could make it impossible to accrete more mass \citep{Packet:1981}.  On the other hand, it may be possible for angular momentum to be efficiently transported outward, allowing accretion to continue \citep{PophamNarayan:1991}.  \citet{Vinciguerra:2020} analysed Be X-ray binaries, in which Be stars were presumably spun up to near-critical rotation by mass transfer from the neutron star's progenitor, and concluded that mass transfer was relatively conservative, i.e., could indeed continue even once the accretor was spun up.

Meanwhile, it is often assumed that accretion onto a compact object cannot exceed the Eddington limit.  Of course, we know of ultra-luminous X-ray binaries hosting neutron stars \citep{Bachetti:2014}, so at least neutron stars can accrete at rates beyond the Eddington limit, presumably by collimating radiation or outflows so that accreting material can bypass the outward pressure.  \citet{King:2003} and \citet{Poutanen:2007} suggested that the Eddington limit can also be surpassed by accreting black holes.  On the other hand, given that thermal timescale mass loss rates from evolved donors can exceed the Eddington limit of compact-object accretors by $\sim 5$ orders of magnitude, even moderately super-Eddington accretion can still be very non-conservative.

How much angular momentum is carried away when mass is lost from the binary during non-conservative mass transfer?  A common assumption in population synthesis codes is that the specific angular momentum of the ejected material is that of the accretor \citep[e.g.,][]{Hurley:2002,COMPAS:2021}.  However, if the accretor is filling its Roche lobe, the material may be lost from the L2 Lagrange point on the side of the accretor away from the donor \citep{MacLeodLoeb:2020gamma}.  This much greater loss of angular momentum reduces the stability of mass transfer, as the orbit shrinks more rapidly.  This would increase the frequency of common-envelope events in binaries.  Figure 2, adopted from \citet{Willcox:2023}, illustrates the sensitivity of mass transfer stability to the specific angular momentum lost from the binary.

\begin{figure}[!t]
\begin{center}
\includegraphics[width=0.8\textwidth]{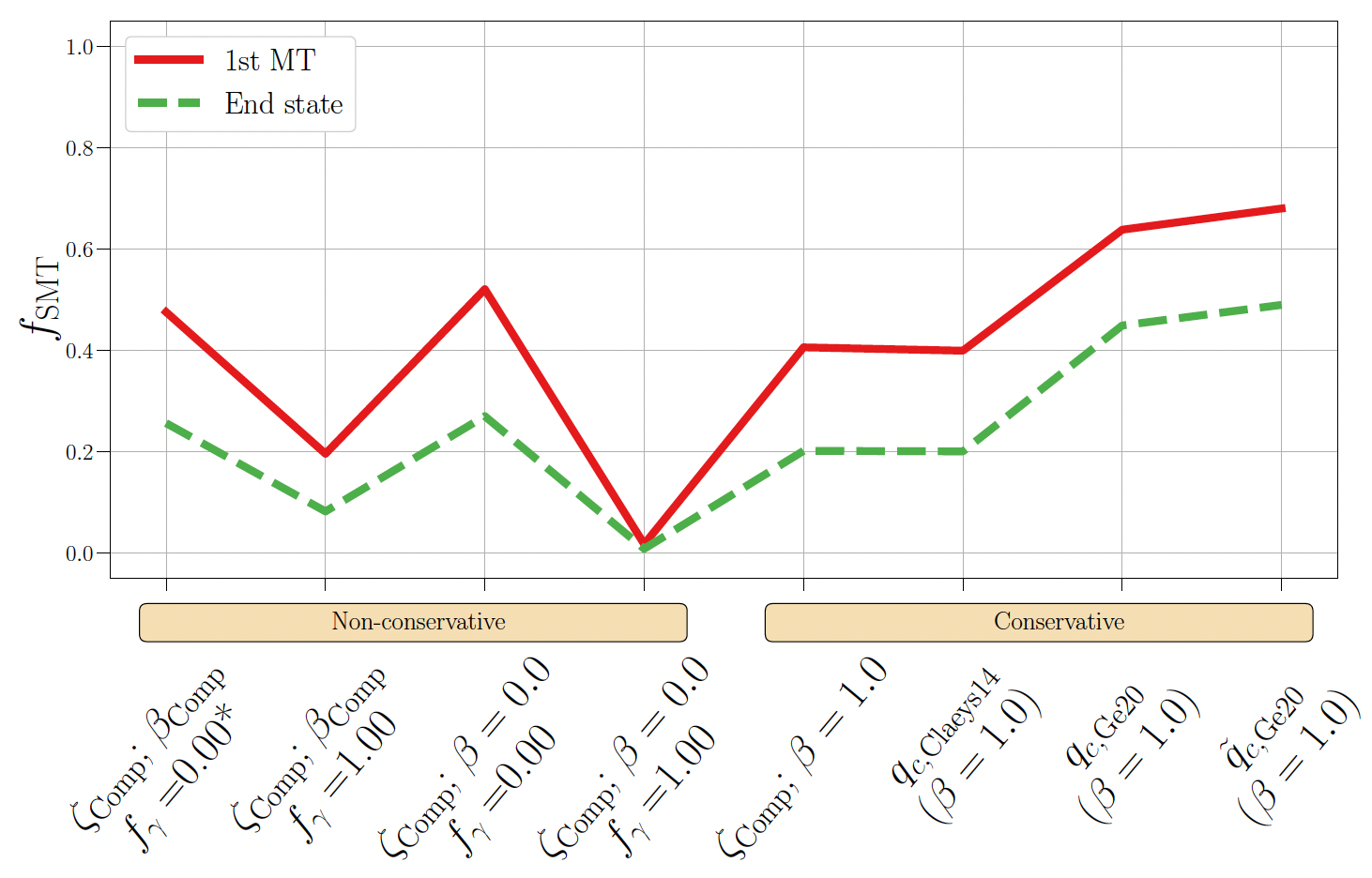}
\end{center}
\caption{The fraction of interacting binaries that experience exclusive stable (as opposed to common-envelope) mass transfer during the first mass transfer event (solid red) and by the end of evolution (dashed green) for a variety of mass transfer assumptions.  $f_\gamma = 1$ corresponds to L2 angular momentum loss, which drastically reduces mass transfer stability relative to $f_\gamma=0$ (specific angular momentum of the accretor).  Figure 5 from \citet{Willcox:2023}.}
\end{figure}

From the above discussion, it seems that common envelopes may be even more common than simplified models suggest.  In the previous section, we mentioned that the standard energy-conserving formalism assumes that energy dissipated through drag goes into heating and ultimately expelling the envelope.  Is this adiabatic assumption reasonable?  It may be for the outer, convective layers of the donor's envelope, which have a flat or outwardly decreasing entropy profile and do re-expand on a dynamical timescale.  However, \citet{Wassink:2021} showed that the radiative intershell between the convective core and convective outer envelope expands on a much longer, thermal timescale.  That means that much of the energy deposited in the radiative intershell can be lost without doing work to expel the envelope.  Therefore, the energy-conserving formalism does not seem appropriate for describing the full common-envelope event.  

Instead, \citet{HiraiMandel:2022} proposed treating the common envelope as a two-stage process, as illustrated in Figure 3.  In our formalism, only the outer, convective layers are treated adiabatically with energy conservation.  Meanwhile, the removal of the radiative intershell below is modelled as any other thermal-timescale, non-conservative mass transfer, through angular momentum conservation.  Overall, this approach means that extreme hardening by two or three orders of magnitude, which is typical for the standard common-envelope formalism, is only seen when there is a significant radiative intershell and the donor's core is appreciably more massive than the accretor, so that the second stage of the common envelope leads to drastic binary hardening.   

\begin{figure}[!t]
\includegraphics[width=0.4\textwidth]{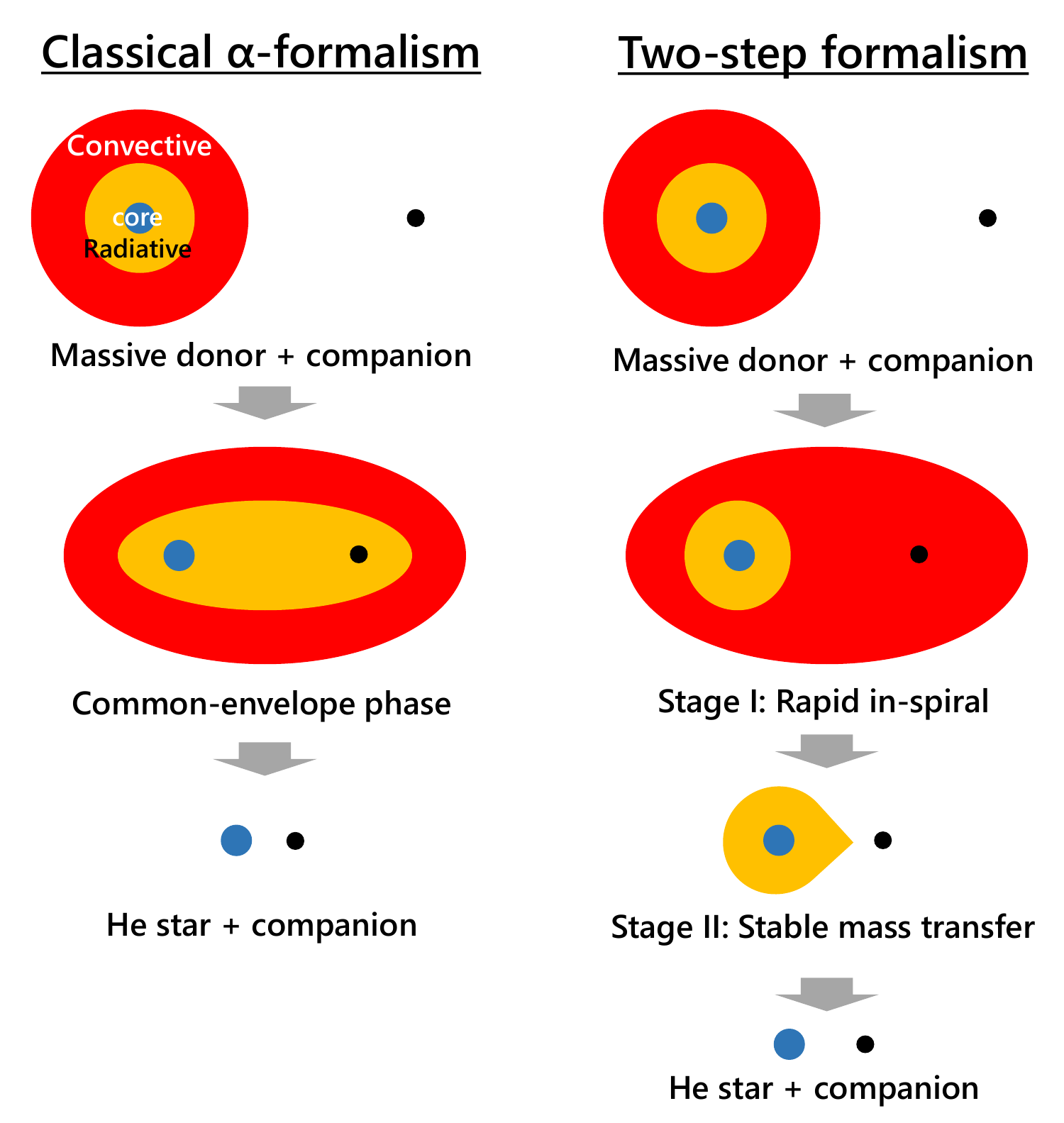} 
\includegraphics[width=0.6\textwidth]{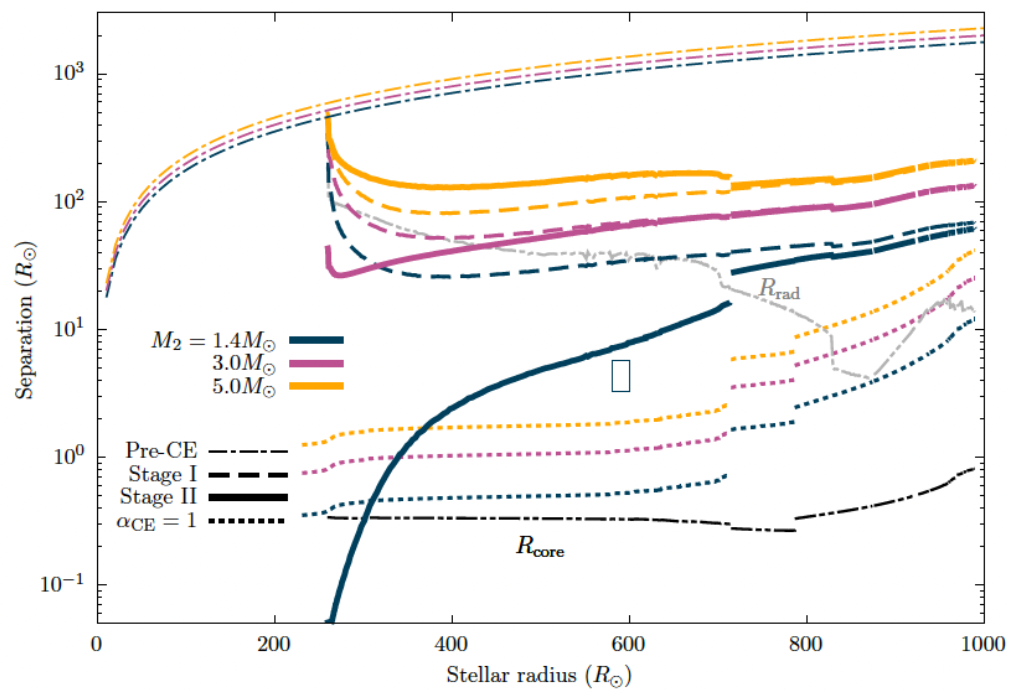}
\caption{Left: a schematic of the energy conserving and two-stage common-envelope formalisms.  Right:  the separations before the onset of mass transfer (dot-dashed curves), after the first-stage outer convective envelope removal (dashed curves) and final separations after the second stage (solid curves) are shown for a  12 M$_\odot$ donor and accretors of given masses.   For comparison, final separations using the energy-conserving formalism with $\alpha=1$ are shown with dotted curves. Figures 2 and 4 from \citet{HiraiMandel:2022}.}
\end{figure}

\citet{Picker:2024} calibrated the masses and binding energies of the convective outer envelopes as functions of stellar mass, evolutionary stage, and metallicity, making it possible to apply the two-stage common-envelope formalism in population synthesis models.  The evolutionary stage is tracked through the effective temperature of the stellar surface.  The convective envelope begins to form when the surface temperature drops to $T_\mathrm{onset}$ and reaches its maximum extent when the temperature drops to $T_\mathrm{min}$.  Analytical fits to the growth of the envelope mass and $T_\mathrm{onset}$ are provided by \citet{Picker:2024} based on MESA stellar models \citep{Paxton:2011}, while $T_\mathrm{min}$ is expected to be taken directly from stellar evolutionary tracks used in a rapid population synthesis code.  We have since realised that this can lead to unphysical behaviour due to mismatches between MESA tracks based on the stellar evolution assumptions used by \citet{Picker:2024} and stellar evolution tracks used in the population synthesis code, e.g., \citet{Hurley:2000} tracks in COMPAS \citep{COMPAS:2021}.  

Here, we report a correction to the \citet{Picker:2024} that sets $T_\mathrm{onset}$ relative to $T_\mathrm{min}$, rather than via Eq.~(6) in that paper, thus avoiding artefacts due to mismatches in single stellar evolution tracks.    As shown in Figure 4, the ratio $T_\mathrm{min}/T_\mathrm{onset}$ is approximately independent of stellar mass over the range $[7, 25]$ M$_\odot$, except for a small deviation at high masses and super-solar metallicities.  We therefore model $T_\mathrm{onset}$ as 
\be
T_\mathrm{onset} = \frac{T_\mathrm{min}}{\mathrm{min} (0.695 - 0.057 \log_{10} Z, 0.95)},
\ee
where $Z$ is the metallicity.

\begin{figure}[!t]
\begin{center}
\includegraphics[width=0.6\textwidth]{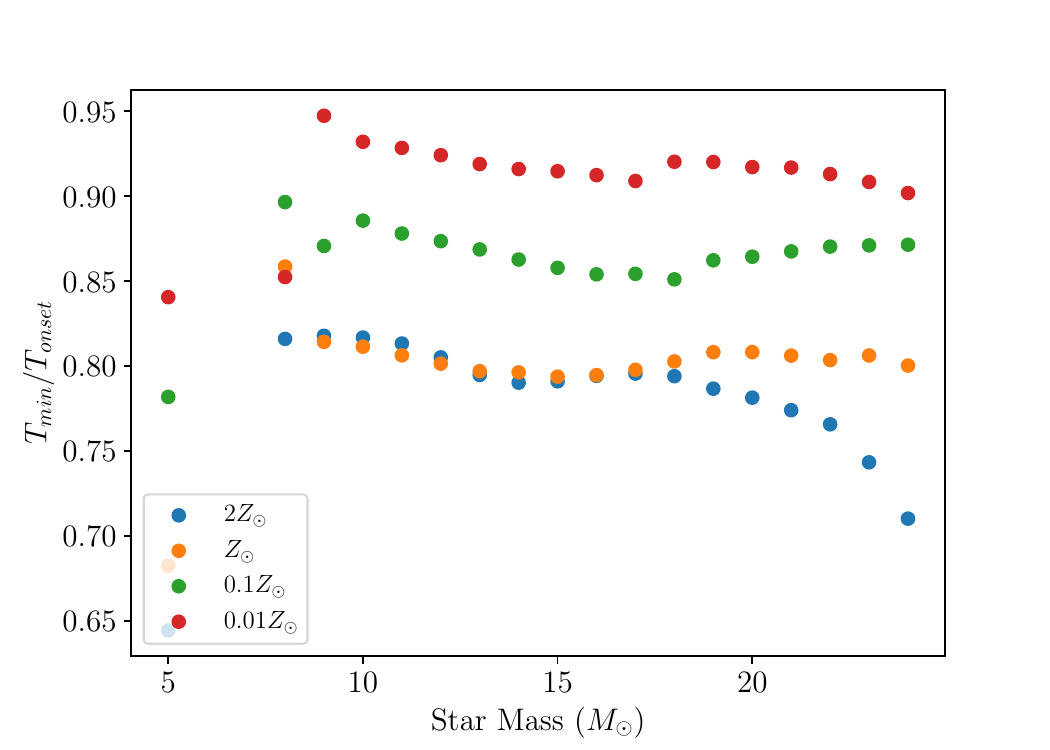} 
\end{center}
\caption{The ratio $T_\mathrm{min}/T_\mathrm{onset}$ between the surface temperature at the maximum extent of the convective envelope and the surface temperature at the onset of convective envelope formation as a function of stellar mass for four choices of metallicity.}
\end{figure}

We expect the two-stage common envelope treatment to have a number of consequences, including significantly decreasing the predicted rate of binary black hole mergers without a similar reduction in the rate of binary neutron stars, relieving some of the tension between population synthesis models and gravitational-wave observations \citep{MandelBroekgaarden:2021}, as well as possibly allowing progenitors of low-mass X-ray binaries to survive the common-envelope phase.

\section{Conclusions}

During the talk, Ilya ran out of time to discuss some of our ongoing work on lightcurves of common envelopes.  Here, we are running out of space.  Briefly, then, let us say that simple population-synthesis models suggest that the Vera Rubin Observatory may image hundreds of luminous red novae associated with common envelopes \citep{Howitt:2020}.  This makes it imperative to improve luminous red nova lightcurve models if we want to use them as observational constraints on common-envelope physics.  Despite very exciting progress with detailed hydrodynamical simulations \citep[e.g.,][]{ChenIvanova:2024}, it remains challenging to accurately extract lightcurves from these.  For example, semi-analytical approaches frequently use Arnett- or Popov-style single-zone models with the addition of recombination energy \citep{Ivanova:2013LRN,MatsumotoMetzger:2022}.  However, unlike supernovae, where there is an instantaneous bomb-like energy injection, the inspiral and inward energy deposition in common envelopes may be slower than the outward energy transport, at least for some configurations \citep[e.g.,][]{Noughani:2024}.  

In summary, then, detailed computational models provide insights into common-envelope physics.  But we also require analytical models to understand and generalise these insights and apply them to stellar populations.  Observationally, a variety of post-common-envelope binaries, as well as upcoming direct observations of ``live'' common-envelope events as luminous red novae with the Vera Rubin Observatory can help us constrain these models.  But we need better analytical understanding to interpret these constraints.  In short, Stan, there is plenty of work here for your retirement!

%-----------------------------------------------------------------------------
% ACKNOWLEDGMENTS:
%
\acknowledgements{We acknowledge support from the Australian Research Council (ARC) Centre of Excellence for Gravitational Wave Discovery (OzGrav), through project number CE230100016.}
%-----------------------------------------------------------------------------

%-----------------------------------------------------------------------------
% BIBLIOGRAPHY:
%
\bibliographystyle{stanfest_bibstyle}
\bibliography{Mandel.bib}
%-----------------------------------------------------------------------------

\end{document}